\documentclass[%
 reprint, 
superscriptaddress,
nofootinbib,
 amsmath,amssymb,
 aps,
pra,  
twoside,
showpacs
]{revtex4-2}

\usepackage{graphicx}% Include figure files
\usepackage{dcolumn}% Align table columns on decimal point
\usepackage{bm}% bold math
\usepackage{ORCIDinREVTeX}

\usepackage[
centering, includefoot,
text={7.1in,10.2in},
total={6.3in,8.75in},
top = 0.8in, bottom = 0.4in,
left=0.62in,right=0.62in
]{geometry}

\makeatletter
\renewcommand\frontmatter@abstractwidth{\dimexpr0.75\textwidth\relax}
\makeatother

\usepackage{siunitx}
\usepackage{physics}
\DeclareSIUnit{\au}{{a.u.}}

\usepackage{nicefrac}
\usepackage{xspace}

\newcommand{\fig}[1]{Fig.\,\ref{#1}}

\newcommand{\labelA}{\textit{A}\xspace}
\newcommand{\labelB}{\textit{B}\xspace}
\newcommand{\labelC}{\textit{C}\xspace}
\newcommand{\labelD}{\textit{D}\xspace}
\renewcommand{\real}{\mathrm{Re}}
\newcommand{\imag}{\mathrm{Im}}
\newcommand{\Rfold}{R_\mathrm{co}}
\newcommand{\thetafold}{\theta_\mathrm{co}}

\newcommand{\thetatwob}{\theta_\star}
\newcommand{\Ip}{\mathcal{I}_\mathrm{p}}
\newcommand{\degree}{^\circ}

\usepackage[
  bookmarks=false,
  colorlinks,
  linkcolor=blue,
  urlcolor=blue,
  citecolor=blue,
  plainpages=false,
  pdfpagelabels,
  final,
  breaklinks=true
]{hyperref}
\hypersetup{
pdftitle={Quantum tunnelling without a barrier}, 
pdfauthor={Anne Weber}
}

\begin{document}

\preprint{APS/123-QED}

\title{\textbf{Quantum tunnelling without a barrier} }% 

\author{Anne Weber}
 \email{Contact author: anne.weber@kcl.ac.uk}
 \orcid{0000-0002-1651-5063}
\author{Margarita Khokhlova}
\orcid{0000-0002-5687-487X}
\author{Emilio Pisanty}
\orcid{0000-0003-0598-8524}

\affiliation{Attosecond Quantum Physics Laboratory, Department of Physics, King’s College London, London WC2R 2LS, United Kingdom}

\date{2 April, 2025}

\begin{abstract}
    Tunnelling is an iconic concept that captures the peculiarity of quantum dynamics, but despite its ubiquity questions remain. 
    We focus on strong-field tunneling, which is vital to all attosecond science. 
    We find an unexpected optical tunnelling event that happens when the instantaneous electric field vanishes and there is no barrier. 
    This event arises from a colour switchover in a strongly polychromatic field. 
    The tunnelling without a barrier reveals the disconnect between the standard intuition built on the picture of a quasi-static barrier, and the nonadiabatic nature of the process.
    \\[-2mm]

    \noindent
    \footnotesize
    Accepted Manuscript for
    \href{%
      https://doi.org/10.1103/PhysRevA.111.043103%
      }{%
      \color[rgb]{0,0,0.55}%
      \textit{Phys.\,Rev.\,A}\,\textbf{111}, 043103 (2025)%
      }, 
    available as %
    \href{%
      https://arxiv.org/abs/2311.14826%
      }{%
      \color[rgb]{0,0,0.55}%
      arXiv:2311.14826%
      }
    under %
    \href{%
      %
      https://creativecommons.org/licenses/by-nc-nd/4.0/%
      }{%
      \color[rgb]{0,0,0.55}%
      CC BY-NC-ND}%
      . 
      Mathematica notebooks to create the figures available 
     \href{%
      https://zenodo.org/records/15104046 
      }{%
      \color[rgb]{0,0,0.55}%
      here%
      }.
    % \\[2mm]

\end{abstract}

\maketitle

%\tableofcontents

\section{Introduction}
    The quantum-mechanical tunnel effect is an emblematic example of the peculiar behaviour of quantum particles, which can `tunnel' through potential-energy barriers that classical physics deems impassable.
    Although the discovery of tunnelling dates back almost a century~\cite{merzbacher2002Early},
    there are still many open questions, 
    such as how to experimentally measure the time the particle spends under the barrier~\cite{buttiker1982traversal, landauer1994Barrier, steinberg1995how, winful2006tunneling, camus2017experimental, ramos2020measurement, schach2024unified},
    tunnelling in composite systems~\cite{balantekin1998quantum, lindberg2023asymmetric, spielmann1994tunneling, benlevy2023simulation}
    and the dynamics inside the barrier~\cite{klaiber2023subbarrier, pisanty2014momentum}.

\begin{figure}[t]
    \centering
    \includegraphics[width=\columnwidth]{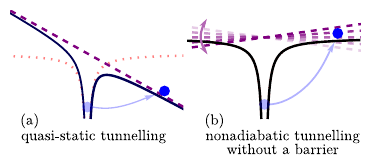}
    \caption{Sketch of field-induced tunnelling, with the laser field, the atomic binding potential and the resulting barrier in the (a) quasi-static approach, and (b) nonadiabatic regime in which the laser field changes during the process.} 
    \label{fig:sketch}
\end{figure}

    In nonlinear optics, tunnelling appears in the context of strong-field physics, in which the illuminating laser field is strong enough to distort an atom's Coulomb potential. 
    The created barrier allows the electron to escape from the atom~\cite{villeneuve2018Attosecond} (see \fig{fig:sketch}(a)).
    In this work, we present such a strong-field ionisation event for an atom subjected to a bichromatic laser field, and report on a nonadiabatic tunnelling event which, counter-intuitively, happens at a time when the instantaneous electric field is~zero (see~\fig{fig:sketch}(b)).

    Ionisation by strong laser pulses is fundamental for the whole area of attosecond science.
    After tunnel ionisation, the electron may be detected directly (known as above-threshold ionisation~\cite{agostini1979FreeFree,becker2002abovethreshold}, ATI), 
    or it may recombine with its parent ion to emit a short burst of radiation (high-harmonic generation~\cite{corkum1993Plasma,Kulander1993}, HHG).
    Both processes provide information on atomic dynamics on its natural timescale~\cite{khokhlova2023Shining},
    and they both initiated by the tunnel step of strong-field ionisation.

    This interaction of intense laser pulses with matter is most commonly described using the strong-field approximation (SFA)~\cite{lewenstein1994Theory,becker2002abovethreshold}. 
    Following the works of Keldysh, Faisal and Reiss~\cite{Keldysh1965, faisal1973Multiple, reiss1980Effect}, and Peremelov et al.~\cite{Perelomov1967}, 
    the ionisation yield in such an oscillating but low-frequency field is derived from the tunnelling rate in a static electric field, and hence, a static barrier. 
    Ionisation can then be described in terms of discrete events which occur when the field is at its maximum.

    In recent years, advances in light generation and measurement techniques allow (and ask) for a more detailed description of the strong-field ionisation process.
    Numerous studies focus on the nonadiabaticity of the tunnelling phenomenon, 
    taking into account the dynamics of the barrier~\cite{yudin2001Nonadiabatic,trabert2021Nonadiabatic, teeny2016Ionization, ni2018Tunneling,barth2011Nonadiabatic}, 
    e.g.\ in the so-called `attoclock' setup~\cite{eckle2008Attosecond, torlina2015Interpreting, zimmermann2016Tunneling, han2019Unifying, han2021Complete, liu2023nonadiabatic, sainadh2020attoclock},    
    as well as the effects of the Coulomb potential~\cite{maxwell2017Coulombcorrected,kelvich2016Coulomb}.

    One of the key tools to investigate nonadiabatic effects are tightly controlled `tailored' polychromatic drivers.
    For example, the perturbative addition of fields with new frequencies is often used as a temporal gate for the ionisation process and to control the harmonic signal~\cite{kneller2022look,shafir2012Resolving, pedatzur2015Attosecond, zhao2013Determination, dahlstrom2011Quantum, mitra2020suppression, ayuso2019synthetic}.
    In the past few years, \textit{strongly} polychromatic drivers, where the two fields are combined with similar amplitudes, have become experimentally feasible and indispensable in attosecond science. 
    They allow for versatile shaping of the harmonic radiation both in intensity and polarisation, and thus offer control over the main features of the emitted attosecond pulses%
    ~\cite{eichmann1995polarization, milosevic2000generation, fleischer2014spin, ivanov2014taking, medisauskas2015generating, chipperfield2009ideal, brixner2004quantum, huang2019realization, eickhoff2021multichromatic, chen2018timeresolved, fang2021optimal}. 
    Generic polychromatic drivers are therefore necessary to build a comprehensive understanding of strong-field processes, starting from their first step~-- tunnelling.

    In this Article we present a tunnel ionisation event that happens in a two-colour strong-field setup at a time when the instantaneous electric field is zero, and hence there is no barrier created.
    This counter-intuitive finding arises within the quantum-orbit picture of the SFA~\cite{kopold2000Quantum} as a generic and topological feature of the colour switchover, which describes the continuous tuning from a monochromatic driving laser field to its second harmonic.
    We suggest this scheme as a simple and essential technique for understanding strong-field physics with strongly polychromatic drivers.

\section{Theoretical Methods}

The ionisation amplitude $\Psi (p)$ for a given final (drift) momentum $p$ can be described by the SFA integral, as shown in~\cite{popruzhenko2014keldysh} and in Appendix \ref{sec:appendixA}. 
In the quantum-orbit picture, we employ the saddle-point method (SPM)~\cite{arfken2013Mathematical, bleistein1975asymptotic} to reduce the integral to a summation of contributions at the saddle points of the semi-classical action~\cite{smirnova2013Multielectron,nayak2019Saddle}
    \begin{equation}
        S(p,t) = \int_{-\infty}^{t} \left[ \Ip + \frac{1}{2} \left(p + A(t')\right)^2 \right] \, \mathrm{d}t' \,.
    \label{eq:action}
    \end{equation}
Here, $\Ip$ is the atomic ionisation potential, the laser is linearly polarised, and its vector potential is \mbox{
{$A(t)= -\int E(t) \,\mathrm{d}t$}. }
(We use atomic units unless otherwise noted.)
The saddle points $t_s$ are determined by
\begin{equation}
    \frac{\partial S(p,t)}{\partial t}\bigg|_{t={t_s}}=0\,, \label{eq:speq}
\end{equation}
they are in general complex numbers, 
correspond to the discrete ionisation events at field maxima from the quasi-static approximation,
and ultimately form a Feynman path integral~\cite{salieres2001feynman} for the ionisation amplitude.
The total ionisation amplitude within one laser cycle can then be written~as
\begin{equation}
    \Psi (p) 
    \approx \Psi_{\mathrm{SPM}} (p) 
    = \sum_s \Psi_{\mathrm{SPM}}^s (p) 
    ,
    \label{eq:IPAsum}
\end{equation}
where $\Psi_{\mathrm{SPM}}^s (p)$ is the ionisation amplitude corresponding to the saddle point $t_s$.
The procedures we describe here have been implemented in Mathematica and are available from \cite{weber_2025_15104046}.

%FIGURE COLOUR SWITCHOVER
\begin{figure}
    \centering
    % FIGURE
        \setlength\tabcolsep{0pt}
        \begin{tabular}[c]{cc}
        \includegraphics[width=0.49\columnwidth]{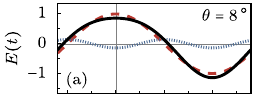} & \includegraphics[width=0.49\columnwidth]{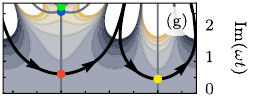} \\
        \includegraphics[width=0.49\columnwidth]{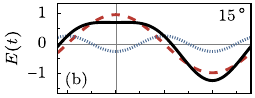} & \includegraphics[width=0.49\columnwidth]{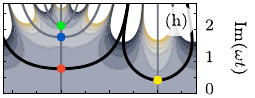} \\
        \includegraphics[width=0.49\columnwidth]{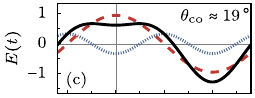} & \includegraphics[width=0.49\columnwidth]{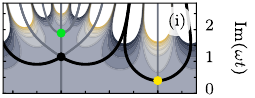} \\
        \includegraphics[width=0.49\columnwidth]{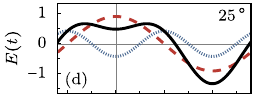} & \includegraphics[width=0.49\columnwidth]{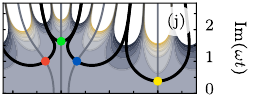} \\
        \includegraphics[width=0.49\columnwidth]{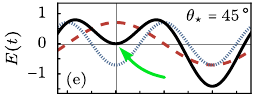} & \includegraphics[width=0.49\columnwidth]{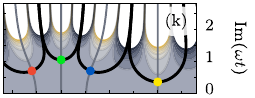} \\
        \includegraphics[width=0.49\columnwidth]{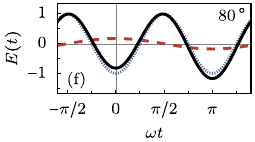} & \includegraphics[width=0.49\columnwidth]{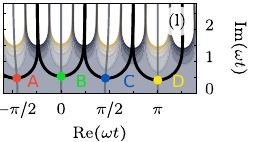}
        \end{tabular}
    % CAPTION
    \caption{Left column (a-f): Total waveform of the bichromatic field \eqref{eq:field} (black solid line) and of its components ($\omega$ field:\ red dashed, $2\omega$ field:\ blue dotted) for the colour switchover performed by increasing the mixing angle $\theta$ from 
    $8\degree$, 
    through $15\degree$, 
    $\thetafold \approx 19 \degree$, 
    $25\degree$, 
    $\thetatwob = 45\degree$ and 
    $80\degree$, respectively.
    The green arrow in panel (e) points at the time when the combined electric field is zero.
    Right column (g-l): $\imag(S(p,t))$ over the complex $\omega t$ plane for the fields presented on the left.
    We use $I_0 = \SI{4e14}{W/cm^2}$, $\omega = \SI{0.057}{\au}$ (so $\lambda = \SI{800}{nm}$), $\Ip = \SI{0.5}{\au}$, such that $\gamma = 0.67$, and $p=0$. 
    Saddle points $\omega t_s$ are highlighted by coloured dots and can be labelled as shown in panel (l).
    Their contour lines for constant $\real(S(p,t))$ are drawn as grey lines, with the resulting integration contour in black.
    }
    \label{fig1}
\end{figure}

\section{The colour switchover}
    We use the saddle-point method to find ionisation events in the colour-switchover scenario. 
    We consider a linearly polarised laser field which is gradually replaced by its second harmonic, such that the resulting electric field can be understood as a two-colour field where we tune the amplitude ratio between the $\omega$ and the $2\omega$ components, while keeping the total intensity $I_0 = E_0^2$ constant. 
    That is, the total electric field is given by
    \begin{equation}
        E(t) = E_1 \cos(\omega t) - E_2 \cos(2 \omega t)    \label{eq:field}
    \end{equation}
    with the amplitudes 
    $E_1 = E_0 \cos\theta$ and 
    $E_2 = E_0 \sin\theta$ defining the
    mixing angle \mbox{$\SI{0}{\degree} \leq \theta \leq \SI{90}{\degree} $}, which corresponds to changing the amplitude ratio \mbox{$R = \nicefrac{E_2}{E_1} = \tan(\theta)$} between $0$ and infinity.

    The colour-switchover scenario, shown in \fig{fig1}, is of special interest because it demonstrates how the number of discrete ionisation events changes as we move from a perturbative second colour to a strongly-polychromatic regime.
    Clearly, for the $\omega$-dominated field there are two ionisation events per cycle of the fundamental, 
    namely at $\omega t=0$ and $\pi$ in panel (a), 
    whereas for the $2 \omega$-dominated field there are four ionisation events per cycle, 
    at $\omega t=-\nicefrac{\pi}{2}$, $0$, $\nicefrac{\pi}{2}$ and $\pi$ in panel (f).
    By examining the contour maps of the imaginary part of the action over complex time (panels (g-l) for the fields in \mbox{(a-f)}), 
    we show that this change in number of contributing saddle points happens surprisingly early within the colour switchover.

    In the contour maps we highlight the saddle points~$t_s$ (coloured dots), their respective level lines (grey solid lines) and the resulting steepest-descent integration path (black solid line) that defines which saddle points contribute to the ionisation amplitude~\eqref{eq:IPAsum} (namely those which are part of the integration path).     
    In the early stage of the colour switchover the integration contour only passes through two saddle points, which we call \labelA (red) and \labelD (yellow).
    The two additional `new' saddle points \labelB (green) and \labelC (blue) come in from high imaginary parts (\fig{fig1}(g,h)), and only start contributing after the coalescence of \labelC and \labelA to one second-order saddle point (black dot in panel (i)), which happens at $\theta = \thetafold $. %$\equiv \arctan(\Rfold)$.
     % \approx \SI{19}{\degree}
    For the rest of the switchover, i.e., for $\theta > \thetafold$ (panels \mbox{(j-l)}), all four saddle points are part of the integration path and hence, must be taken into account when calculating the ionisation amplitude.

    Let us focus on the saddle point \labelB, for which \mbox{$\real(\omega t_s) = 0$}.
    We find that this saddle point represents a contributing ionisation event even before the electric field has changed sign ($E(0) > 0$, as in \fig{fig1}(d)) and therefore implies tunnelling \textit{uphill}. 
    More importantly however, and as advertised, this tunnel ionisation event also contributes to the total ionisation amplitude at $\thetatwob=\SI{45}{\degree}$
    (\fig{fig1}(k), saddle point \labelB at $\omega t_s \approx 0 + 1.1\,\mathrm{i}$),
    when the electric field is zero 
    ($E(0) = 0$, green arrow in \fig{fig1}(e)) 
    and there is in fact no barrier formed.

%%% [EXPLANATION IN THE COMPLEX PLANE]
    This finding is deeply surprising when thinking of strong-field ionisation in terms of the well-established quasi-static intuition~\cite{Keldysh1965, faisal1973Multiple, reiss1980Effect, Perelomov1967}.
    One possible explanation is to account for the nonadiabaticity of the process using the complex-time model explained in~\cite{smirnova2013Multielectron}.
    Therein the complex-valued saddle point $t_s$ marks the moment the electron enters the barrier,
    $\imag(t_s)$ is understood as the time it spends under the barrier, and $\real(t_s)$ is the moment it appears in the continuum (read: exits the barrier).
    As, in our case, the electric field is non-zero during that tunnelling time, i.e., for the time between $t_s$ and $\real(t_s)$, one could argue that there is a barrier formed which allows the electron to tunnel.
    However, the dynamics in complex-valued time are far from unambiguous.

It is worth emphasizing that the tunnel-ionisation dynamics hold in essentially identical form regardless of the value of the drift momentum $p$, and the only special feature at $p=0$ is the exact saddle-point coalescence of \fig{fig1}(i).
In particular, for any given $p$ we can always find an ionisation event happening at a time when the electric field is zero.

\begin{figure}[t]
    \centering
    \includegraphics[width=\columnwidth]{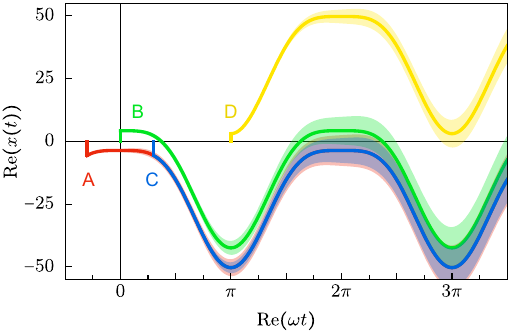}
    \caption{Semi-classical trajectories~\eqref{eq:trajectory} for the field with the four labelled ionisation events shown in \fig{fig1}(e) and (k) respectively.
    The semi-transparent bands show the trajectories for solutions with small drift momentum, \mbox{$|p|< \SI{0.05}{\au}$}}
    \label{fig:trajectories}
\end{figure}

\section{Electron trajectories}
    To further expand on the intuitive understanding of the ionisation events, we show the electron's trajectory in the electric field, given by
    \begin{equation}
      x(t) = \int_{t_s}^t \left( p + A(t') \right) \,\mathrm{d}t' \, . \label{eq:trajectory}
    \end{equation}
    The temporal integration here follows a two-legged contour in the complex plane~\cite{pisanty2016Slalom,torlina2012Timedependent,smirnova2013Multielectron}, 
    starting from the complex-valued saddle point $t = t_s$ downwards to the real axis, 
    and then from $t = \real(t_s)$ along the real axis.
    In \fig{fig:trajectories} we show the resulting semi-classical trajectories for all four ionisation events shown in \fig{fig1}(k).
    In general, the position at the tunnel exit 
    $x_\mathrm{exit}=\real\big[x(\real(t_s))\big]$ 
    is non-zero for all trajectories.
    Upon their appearance in the continuum, trajectories \labelA, \labelB and \labelC are driven away from the core, return, and then remain in the core's vicinity for a significant fraction of the cycle (around $\omega t \approx 2\pi$).

    These trajectories are therefore highly susceptible to Coulomb effects,
    which would weaken their contribution to the measurable ionisation yield.
    However, the ionisation event at zero field is a stable feature of the colour switchover and remains even for a wide range of phase shifts between the two constituent fields. 
    It is therefore possible to adjust the colour-switchover configuration to produce zero-field tunnelling events at nonzero momenta whose trajectories spend less time around the core and for which Coulomb forces are minimised.
    Independently of Coulomb-effect considerations, the appearance of zero-field tunnelling events within the SFA framework is interesting in its own right.

\section{Contributions to the total ionisation amplitude}
\begin{figure}[t]
    \centering
    \begin{tabular}[c]{c}
    \includegraphics[width=\columnwidth]{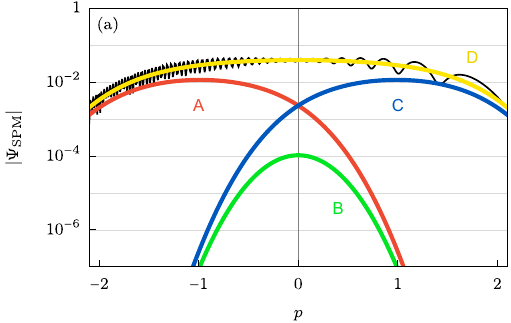} \\ 
    \includegraphics[width=\columnwidth]{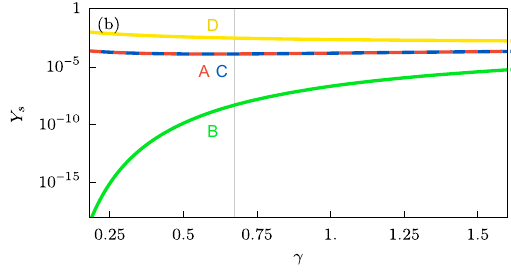}
    \end{tabular}
    \caption{(a) Magnitude of the spectral ionisation amplitude $|\Psi_\mathrm{SPM}(p)|$ 
    (black), and the contribution $|\Psi_\mathrm{SPM}^s(p)|$ of each of the four ionisation events shown in \fig{fig1}(k).
    For $p=0$, the contribution \labelB (green) stems from the ionisation event at zero field.
    (b) Scaling of the total ionisation probability per orbit, $Y_s$, for a field shaped like the one shown in \fig{fig1}(e), for a range of Keldysh parameters $\gamma = 4 \omega \sqrt{\nicefrac{\Ip}{5 I_0}}$, as a function of $\omega$ for constant $\Ip$ and $I_0$, over the wavelength range $\SI{330}{nm} \leq \lambda \leq \SI{3000}{nm}$. %336.859
    The configuration used for Fig.\,\ref{fig1}, Fig.\,\ref{fig:trajectories} and panel (a) yields $\gamma = 0.67$ (grey vertical line).
    Note that the total contributions \eqref{eq:integratedcontribution} of \labelA and \labelC are equal. 
    }
    \label{fig:spectrumscaling}
\end{figure}

    Let us now turn to the contribution of each ionisation event to the total ionisation amplitude~\eqref{eq:IPAsum}, where we are especially interested in the contribution of the zero-field ionisation event \labelB.
    Hence, in \fig{fig:spectrumscaling}(a) we show the spectrum $| \Psi_{\mathrm{SPM}} (p) |$ (black) for the field shown in \fig{fig1}(e), with the contributions of each ionisation event.
    We find that the contribution of our peculiar ionisation event is small compared to that of \labelA and \labelC, and particularly \labelD. 
    In fact, for the majority of the colour switchover (field shapes as in Figs.~\ref{fig1}(c-f)) the spectrum is clearly dominated by the contribution of \labelD, such that the contribution of orbit \labelB is hidden below the others and does not have an immediately observable effect.
    This becomes obvious when we recall that the ionisation for orbit \labelD happens when the field amplitude is largest (around $\omega t \approx \pi$), and that the instantaneous electric field enters exponentially into the ionisation amplitude via the action~\eqref{eq:action}.

% [SCALING BEHAVIOUR]
    Because the theoretical SFA framework is tightly linked to, and is often benchmarked against, the assumption of a quasi-static barrier, we explore how the relevance of the zero-field tunnelling event changes as we approach the adiabatic limit.
    Firstly, it is essential to mention that the instantaneous electric field is zero at $t=0$ for amplitude ratio $R=1$, and hosts a contributing saddle point there, independently of the Keldysh parameter (apart from the static limit case of $\gamma =0$) as shown in Appendix B.
    Building on that, in \fig{fig:spectrumscaling}(b) we present each orbit's total contribution to the spectrum,
    \begin{equation}
        Y_s = \int | \Psi_{\mathrm{SPM}}^s (p) |^2
        \, \mathrm{d} p \,, \label{eq:integratedcontribution}
    \end{equation}
    for various Keldysh adiabaticity parameters $\gamma = \sqrt{\nicefrac{\Ip}{2 U_p}}$ and the ponderomotive energy , with $U_p$ the ponderomotive energy.
    From the quasi-static limit ($\gamma \ll 1$) to the multi-photon ionisation regime ($\gamma > 1$), orbit \labelD dominates the spectrum, followed by equal contributions by \labelA and \labelC, all of which show little variation across the range of Keldysh parameters.
    In contrast, we find that the contribution of orbit \labelB decreases as we approach the quasi-static regime. 
    This means that the tunnelling event at zero field loses significance for the total spectrum as we move towards the adiabatic limit,
    reassuringly reconciling our results with the quasi-static intuition.
    Nevertheless, we emphasise that the experimentally-relevant parameters used above result in $\gamma = 0.67$. %0.673718 
    Thus, within the realms of typical configurations like ours, the contribution of orbit \labelB must not be~ignored.
    Here it is worth remarking that the colour switchover can also be performed between $\omega$ and $3 \omega$.
    That produces a symmetric time dependence of the field without the large maximum at $\omega t \approx \pi$, and narrows the dynamics down to only orbits \labelA, \labelB and~\labelC.
    
\section{Outlook and Conclusions}
    Looking forward, the counter-intuitive nature of the tunnelling without a barrier makes it highly desirable to work towards experimental realisation of this phenomenon and, as a prerequisite, to identify an observable for which the tunnel ionisation event at zero field produces detectable signatures.
    Then, naturally, we aim for a comparison both with numerical simulations of the time-dependent Schrödinger equation (TDSE) and with experimental~data.
    TDSE simulations of this configuration, in particular, should be able to clarify to what extent the effects of the Coulomb potential, as well as the possibility of rescattering, play a role in the contribution of zero-field tunnelling events to observable signatures.
    Those observable signatures could be drawn, for example, from two-dimensional holographic interference patterns in photoelectron momentum spectra.

    Lastly, performing a colour switchover in the context of HHG instead of ATI 
    promises to show similar features to the ones described here,
    but it also offers the possibility to emphasise the role of the zero-field tunnelling event through the use of spatial interference and macroscopic phase-matching effects.

%%% [CONCLUSION]
    In conclusion, here we present strong-field tunnel ionisation events in the colour switchover from a laser field to its second harmonic.
    This scheme generically covers all possible two-colour fields~\eqref{eq:field}, for which the ionisation process is well described by our SFA framework, and is intuitively understood within quasi-static assumptions.
    In the case of equal-amplitude mixing of the two constituent fields (as in \fig{fig1}(e)), we find a nonadiabatic tunnel ionisation event that happens at a time when the instantaneous electric field is zero, and hence there is no barrier, thereby challenging that intuitive picture of the process.
    The existence of the event is a topological feature of the two-colour field~\eqref{eq:field} at equal amplitudes, and has a non-zero contribution to the total spectrum from the quasi-static limit through the multiphoton regime. 
    We find that the event has a relatively small contribution to the spectral ionisation amplitude, but we expect it to play a detectable role in the heterodyne diffraction patterns of the photoelectron momentum distributions and to initiate contributing trajectories for the generation of high-order harmonics.
    These questions thus invite the search for an observable that provides experimentally measurable signatures of the tunnelling at zero field.

\begin{acknowledgments}
AW and EP acknowledge Royal Society funding under URF\textbackslash R1\textbackslash 211390 and RF\textbackslash ERE\textbackslash 210255. MK acknowledges Royal Society funding under URF\textbackslash R1\textbackslash 231460.
\end{acknowledgments}

\appendix

\section{Ionisation Amplitude}\label{sec:appendixA}
    We consider ATI within the SFA framework.
    The ionisation amplitude for a given final (drift) momentum $p$ is therein described by the integral
    \cite{popruzhenko2014keldysh}
    \begin{equation}
        \Psi (p) = \int_{-\infty}^{\infty} \mathcal{P}\left(p + A(t)\right) \,
        \mathrm{e}^{-\mathrm{i}S(p,t)} \,\mathrm{d}t \label{eq:IPAintegral}
    \end{equation}
    where $S(p,t)$ is the semi-classical action %~\eqref{eq:action}.
    as shown in the main text of this Article,
    {${\mathcal{P}(k) = \mathrm{i} \left(\Ip + \nicefrac{k^2}{2} \right) \varphi_0(k)}$}
    is a slowly-varying prefactor,
    $\Ip$ is the ionisation potential, 
    and 
    $\varphi_0$ is the momentum representation of the ground-state wavefunction.
    For atomic targets we assume a short-range potential such that
     $\mathcal{P}(k) = \nicefrac{\mathrm{i}}{\sqrt{\pi}}\left(2 \Ip \right)^{\frac{1}{4}}$.  
    Applying the saddle-point method (SPM) to the SFA integral \eqref{eq:IPAintegral} results in a summation over discrete ionisation events $t_s$:    
        \begin{align}
        \Psi_{\mathrm{SPM}} (p) 
        &\approx  
        \sum_s \Psi_{\mathrm{SPM}}^s (p) \nonumber \\
        &=
        \sum_s 
        \sqrt{\frac{2 \pi}{\mathrm{i} S''}} \, 
        \mathcal{P} \left(p + A(t_s)\right) \,
        \mathrm{e}^{-\mathrm{i}S(p,t_s)}\,,  %\label{eq:IPAsum}
        \end{align}
    where the $S''$ is the second derivative of the action evaluated at the saddle point $t_s$.
    An often neglected step of applying the saddle-point method (also called method of steepest descents) is that the summation does not include \emph{all} saddle points~$t_s$ of the action, but only those which are part of a valid integration path~\cite{bleistein1975asymptotic}.
    In fact, there typically exist numerous solutions to the saddle-point equation \ref{eq:speq}, from which we need to select the ones which are contained in the steepest-descent route.
    For the considered colour-switchover scenario, those steepest-descent integration contours are shown in Fig.\,1 (g-l) of the main text.

\section{Scaling of the saddle-point coalescence}

%% [COALESCENCE]
    Strictly speaking, whenever two saddle points coalesce, or are in close proximity, the saddle-point method breaks down.
    Approximating the SFA integral then requires higher-order methods, such as uniform expansions~\cite{bleistein1975asymptotic}, which for our case are complicated by the presence of the third saddle point \labelB.
    A more detailed exploration follows in an upcoming publication. 
    Regardless of that, the coalescence point~$R = \Rfold$ indubitably marks the point from which on the `new' saddle points \labelB and \labelC (\labelB eventually corresponding to the ionisation event at zero field) enter the integration contour.
    That is, for a given configuration, the saddle point \labelB (as well as \labelC) needs to be taken into account for all amplitude ratios $R > \Rfold$.

\begin{figure}[ht]
        \centering
        \includegraphics[width=\columnwidth]{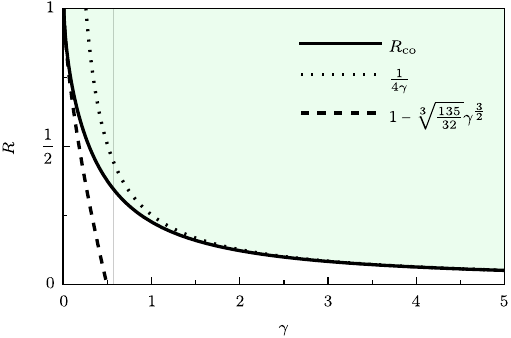}
        \caption{Scaling of the amplitude ratio $\Rfold$ at which the saddle-point coalescence happens, over a range of Keldysh parameters $\gamma$. %, as in \fig{fig:spectrumscaling}(b).
        The shaded region shows the parameters for which saddles \labelB and \labelC contribute to the integration contour.
        In dashed lines, asymptotes for the behaviour in the $\gamma \ll 1$ and $\gamma >1$ regime are shown.
        The grey line marks the configuration used in Fig.\,2 of the main text, %\fig{fig1},
        for which $\gamma = 0.67$ and the coalescence happens at $\Rfold \approx 0.36$, corresponding to $\thetafold \approx \SI{19}{\degree}$.
        }
        \label{fig5}
    \end{figure}

%%% [SCALING OF THE COALESCENCE] 
    As mentioned in the main text, the electric field is zero for equal constituent field amplitudes ($R=1$) independently of the Keldysh parameter.
    To verify that the tunnelling event without a barrier remains a contributor to the total ionisation amplitude for a change of driving laser field parameters we need to show that saddle point \labelB is already part of the integration contour at $R=1$, even if the wavelength changes.    
    In \fig{fig5} we therefore show that the coalescence amplitude ratio $\Rfold$ decreases monotonically with the Keldysh parameter (solid line). 
    The dependence can be well approximated by 
    the asymptotes $1 - \sqrt[3]{\nicefrac{135}{32}}\, \gamma^{\frac{3}{2}}$ in 
    the small-wavelength limit 
    and by $\nicefrac{1}{4 \gamma}$ in the long-wavelength limit (dashed and dotted lines respectively).
    Most importantly, \fig{fig5} shows that $\Rfold < 1$ for all $\gamma > 0$ (shaded region), and 
    the saddle point \labelB at the equal-amplitudes line $R=1$ is always part of the integration contour.
    We hence conclude that the ionisation event at zero field is a topologically stable feature of the strongly bichromatic driving field.

\vfill

% \nocite{*}

\bibliography{references}% Produces the bibliography via BibTeX.

\end{document}